\documentstyle[prl,aps,twocolumn,epsfig]{revtex}

\begin{document}

\preprint{}
\draft

\title {Fidelity Decay Saturation Level for Initial Eigenstates}
\author{Yaakov S. Weinstein$^\dag$, Joseph V. Emerson$^\dag$, Seth, Lloyd,$^*$
David G. Cory$^{\dag\sharp}$} 

\address{$^\dag$Massachusetts Institute of Technology, Department of Nuclear 
Engineering, Cambridge, MA 02139 \\
$^*$d'Arbeloff Laboratory for Information Systems and Technology,
Massachusetts Institute of Technology, Department of Mechanical Engineering, 
Cambridge, MA 02139 \\
$^\sharp$Author to whom correspondence should be addressed}

\maketitle

\begin{abstract}
We show that the fidelity decay between an initial eigenstate state evolved 
under a unitary chaotic operator and the same eigenstate evolved under a
 perturbed operator saturates well
before the $1/N$ limit, where $N$ is the size of 
the Hilbert space, expected for a generic initial state. 
We provide a theoretical argument and numerical evidence 
that, for intermediate perturbation strengths,
the saturation level depends quadratically on the perturbation strength. 
\\
PACS numbers 
\\
\end{abstract}

Over the past twenty years different phenomenon found in quantum systems
that have chaotic classical analogs have been suggested as appropriate 
signatures of quantum chaos 
\cite{Berry}\cite{BGS}\cite{Haake}\cite{P1}\cite{SC}\cite{ZP1}. 
Peres \cite{P1} conjectured that the initial rate and behavior of a 
system's fidelity decay due to a small perturbation in the Hamiltonian 
may provide an appropriate signature of quantum chaos. 
This signature provides an analog to the 
sensitivity to initial conditions which characterizes classical
chaos but, as a consequence of strictly unitary evolution, 
cannot emerge in quantum systems. 
Recent insights \cite{Prosen}\cite{Jacq}\cite{J} 
have lead to a more detailed understanding of this signature. 
 
For a unitary map, $U$, the fidelity compares the evolution of an 
initial state under unperturbed and perturbed dynamics. 
The fidelity is given by 
\begin{equation}
F(n) = |\langle\psi_i|(U^{\dagger})^n(U_pU)^n|\psi_i\rangle|^2
\end{equation} 
where $U_p = exp(-i\delta V)$ is the perturbation operator of strength 
$\delta$, and $\psi_i$ is the initial state of the system. 
The fidelity decay behavior depends not only on whether the map is chaotic 
but also on the initial state of the system and the 
strength of the perturbation. For chaotic systems, the fidelity eventually 
approaches an asymptotic level. Here, we focus on the characteristics of
this asymptotic level by studying  
\begin{equation}
F_{\infty} = \lim_{n\rightarrow\infty}\frac{1}{n} \sum^{n}_{1}F(n)dn.
\end{equation}
For initial random states the fidelity saturates at $1/N$ \cite{Prosen}, 
as we show below./ However, for eigenstates $F_{\infty}$, 
the saturation level, in much larger and depends 
sensitively on the perturbation strength, $\delta$. 
The study of initial eigenstate fidelity decay is of particular interest since
it is equivalent to the survival probability of a system eigenstate under the
influence of a perturbation. Below we provide theoretical 
arguments showing a region where $F_{\infty}$ depends quadratically on the
perturbation strength. We also test this prediction numerically on 
quantum chaotic maps. 

For chaotic systems and weak perturbation strengths the initial 
fidelity decay is Gaussian, as expected from perturbation 
theory and random matrix theory\cite{Tom}. For stronger perturbations the 
fidelity decay is exponential. 
The regime of exponential decay, known as 
the Fermi Golden Rule (FGR) regime \cite{Jacq}, is reached when 
$\sigma$, a typical off-diagonal matrix element of perturbation 
Hamiltonian expressed in the ordered eigenbasis of the system 
Hamiltonian, is greater then the average system level spacing, $\Delta$. 
It has been shown that for some perturbations the rate of the the exponential 
decay increases as $\delta^2$, 
until saturating at a rate given by the corresponding classical 
system's Lyapunov exponent \cite{Jala}\cite{Jacq}\cite{Cas}, 
or the bandwidth of the system Hamiltonian \cite{Jacq}. 
  
Jacquod and coworkers \cite{Jacq} showed that the fidelity 
decay in the FGR regime
is related to the local density of states (LDOS) for initial 
eigenstates. We define eigenvectors and eigenangles for the unperturbed 
operator $U|v_j\rangle = exp(-i\phi_j)|v_j\rangle$, and the perturbed operator
$U_pU|v_k'\rangle = exp(-i\phi_k')|v_k'\rangle$. The LDOS is the spectral 
density of the original system under transition rules given by the 
perturbation. Hence, it is  as a measure 
of the overlap between perturbed and unperturbed states separated by an angle
$(\phi_j - \phi_k')$
\begin{equation}
\eta(\phi_j - \phi_k') = |\langle v_j|v_k'\rangle|^2.
\end{equation}
For an initial eigenstate of $U$, the fidelity decay is the Fourier
transform of the LDOS 
\begin{equation}
F(n) = \sum_m\eta (\phi_j-\phi_k')exp(-i(\phi_j-\phi_k')n).
\end{equation}
Previous studies suggest that the LDOS of a complex system in the regime of 
strong perturbation is Lorentzian\cite{Wigner,FCIC,Jacq2}
\begin{equation}
\eta (\phi_j-\phi_k') = \frac{\Gamma}{(\phi_j-\phi_k')^2+(\Gamma/2)^2}
\end{equation}
with a width of $\Gamma = 2\pi\sigma^2/\Delta$ where $\sigma$ is a typical 
off diagonal element of the perturbation operator. Thus, using the Fourier 
transform relation, the initial fidelity decay is exponential with a rate 
of $\Gamma$
\begin{equation}
F(n) = exp(-\Gamma n).
\end{equation}

$\Gamma$ can be rewritten in terms of perturbation strength as follows: 
$\sigma=\sqrt{\delta^2\overline{V_{mn}^2}}$ 
where $\overline{V_{mn}^2}$ is the second moment of the matrix elements 
$V_{mn}$. $\overline{V_{mn}^2}$ may be estimated by noting that for 
chaotic systems the eigenvectors are random,    
and, therefore, $\overline{V_{mn}^2}  = \overline{\lambda^2}/N$ \cite{J} where 
$\overline{\lambda^2} = N^{-1}\sum_{i=1}^N\lambda^2_i$ is the variance of the
eigenvalues of $V$. The average level spacing, $\Delta$, is equal to $2\pi/N$.
The rate of the exponential decay, $\Gamma$ can now be evaluated as  
$\Gamma = \delta^2\overline{\lambda^2}$. 

We now turn to the study of the saturation level of the fidelity decay, 
$F_{\infty}$. After a certain amount of time we would expect the initial
state to become evenly spread out over a complete set of states. This
implies that $F_{\infty}$ should be of order $1/N$. 
$F_{\infty}$ should also be independent 
of the perturbation strength. A weaker perturbation simply leads to a 
longer period of time until the saturation level is reached, but the 
saturation level should remain unchanged.
For initial states that are eigenstates of the unperturbed system, however, 
$F_{\infty}$ depends on the perturbation strength. Weaker perturbations, even
in the FGR regime, lead to saturation levels significantly higher then
$1/N$.

Prosen \cite{Prosen} has noted that for initial eigenstates
$F_{\infty} \rightarrow 1$ in the limit of weak perturbation and 
$F_{\infty} \rightarrow (4-\beta)/N$ for strong perturbation where 
$\beta = 1$ for maps
with circular orthogonal ensemble (COE) properties and $\beta = 2$ for maps 
with circular unitary ensemble (CUE) properties. Here, 
we provide a theoretical argument and numerical evidence for
a quadratic behavior for $F_{\infty}$ of initial eigenstates versus 
perturbation strength for perturbation strengths between these two extremes.  

To evaluate the dependence of the saturation level on the 
perturbation strength let us start by expressing the 
fidelity for an initial eigenstate, $|v_m\rangle$, as
\begin{equation}
F(n) = |\langle v_m|\sum_l a_{lm}e^{-in(\phi'_l-\phi_m)}|v_m\rangle|^2
\end{equation}
where $a_{lm} = \langle v_l'|v_m\rangle$. The above equation can be 
separated into a time independent term plus a time dependent term 
\begin{equation}
F(n) = \sum_l |a_{lm}|^4 + 
\sum_{lk}|a_{lm}|^2 |a_{km}|^2\cos[(\phi_l'-\phi_k')n].
\end{equation}
The time average of the second term goes to zero while the first term 
determines $F_{\infty}$ as an inverse participation ratio
of the overlap between perturbed and unperturbed eigenvectors\cite{Prosen}.
In other words, the fidelity saturation level is simply the sum of the squared 
elements of the LDOS. 

Using equation (8), we recover the $\simeq 1/N$ saturation level of Prosen  
in the limit of strong perturbation.
An extremely strong perturbation could cause the initial 
state (though an eigenstate of the system dynamics) to become evenly spread 
over all eigenstates of the system, such that 
$|\langle v_l'|v_m\rangle| \simeq 1/\sqrt{N}$ for all $|v_l'\rangle|$. 
$F_{\infty}$ would then be $\sum_l |a_{lm}|^4 \simeq 1/N$. 
For weaker perturbations the saturation 
level will depend on the number of contributing eigenvectors, $|v_l'\rangle$
and the coefficients $a_{lm}$. This is equivalent to the width of the LDOS 
under the particular perturbation.

Hence, to estimate $F_{\infty}$ for intermediate strengths 
we must have an idea of the number of 
contributing perturbed operator eigenvectors $|v_l'\rangle|$ to the initial 
eigenstate, $|v_m\rangle$. This can be estimated 
by the width of the LDOS, $|a_{lm}|^2$. 
We assume all eigenvectors within the width $\Gamma$ of the approximate 
Lorentzian shaped LDOS to have equal weight. With this approximation, 
\begin{equation}
F_{\infty} \propto 1/(\Gamma N) = 1/(\delta^2\overline{\lambda^2} N). 
\end{equation}
Thus, we expect a quadratic dependence of $F_{\infty}$ on the 
perturbation strength
in the FGR regime until the saturation level reaches $O(1/N)$.

A similar analysis for an initial random state, 
$|\psi_i\rangle = \sum_m c_m|v_m\rangle$, shows that 
$F_{\infty} \simeq 1/N$ for all perturbation strengths. 
For random states the fidelity can be written as 
\begin{equation}
F(n) = |\sum_{mlj}c_m^* c_j a_{lj} a_{lm}^* e^{-in(\phi_l'-\phi_m)}|^2.
\end{equation}  
Once again, the right hand side can be divided to a time dependent term
\begin{equation}
\frac{1}{N^2}\sum_{mlj}\sum_{m'l'j'}a_{lj}a_{l'j'}^* a_{lm}^* a_{l'm'}
\cos[(\phi_l'-\phi_{l'}'+\phi_{m'}-\phi_m)n]),
\end{equation}
which vanishes under time average, 
and time independent term
\begin{equation}
F_{\infty} = \frac{1}{N^2}\sum_{mlj}|a_{lj}|^2 |a_{lm}|^2
\end{equation} 
where, in the above equations, $|c_j|^2 \simeq 1/N$ for a random state. 
For any non-zero perturbation strength, 
the time average of time dependent term will go to zero. 
The time independent term is easily seen to be approximately 
$1/N$ in the limits of weak and strong perturbation. For intermediate 
perturbation strengths we can estimate the contribution of the 
time independent term by analyzing the LDOS. Again, we 
approximate the Lorentzian LDOS with a rectangle of width $\Gamma$ and 
height $1/\Gamma$. Contributions to the sum will be non-zero only if the
$j$th and $m$th eigenvectors are a distance of less than $\Gamma/2$ from 
the $l$th perturbed eigenvector. Hence, for each of the $N$ values of 
$l$ there will be $\Gamma$ terms $|a_{lj}|^2$ and $\Gamma$ terms $|a_{lm}|^2$ 
each of magnitude $1/\Gamma$. The value of the time independent term 
is thus $1/N$.

The above predictions were first tested on random circular
unitary ensemble (CUE) maps. Random matrix theory 
predicts the behavior of the fidelity decay in both the Gaussian\cite{Tom} and 
FGR\cite{Jacq3} perturbation strength regimes. The use of a random matrix as
the evolution operator to study dynamical aspects of quantum chaos has been 
done in \cite{J}. 

We assume that our system is composed of a collection of two-level subsystems 
or qubits. The perturbation used is a $z$-rotation of all of these qubits
through an angle $\delta$ 
\begin{equation}
U_p = \prod^{n_q}_{j=1} e^{-i\delta\sigma^j_z/2}
\end{equation}
where $n_q = log_2 N$ is the number of qubits in the system. In the context of
quantum information processing, this perturbation corresponds to an
error in the phase of all the quantum bits in a quantum information processor.
We note that this perturbation also arises in quantum control studies as a 
model of coherent far-field errors \cite{Viola}. For this perturbation, CUE 
maps exhibit exponential fidelity decay and a Lorentzian 
shaped LDOS \cite{J} as shown in the insets of figure 1. 


Figure 1 shows $F_{\infty}$ versus perturbation strength for CUE maps, 
using initial eigenstates of the CUE matrix. We see that below the FGR regime
there is very little decay while in the limit of strong perturbation
$F_{\infty} = 2/N$ as expected for CUE maps. Between these we see a power law 
decrease of $F_{\infty}$ with increased perturbation strength. 
Since the LDOS is Lorentzian the discrepancy seen in figure 1 must be 
due to the approximation made by replacing the Lorentzian LDOS with a 
rectangle of width $\Gamma$. The actual slope of the data is between 
1.8 and 1.9. The data is compared to  
$F_{\infty}  = C_{CUE}/(\delta^2\overline{\lambda^2} N)$, 
where the proportionality constant, 
$C_{CUE} = 3.6$ is chosen to best fit the data. 

\begin{figure}
\epsfig{file=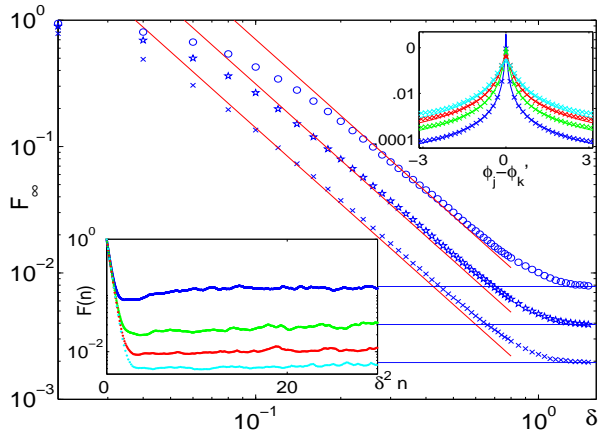, height=5.8cm, width=8cm}
\label{d}
\caption {Saturation level versus perturbation strength for initial eigenstates
of a random CUE map of dimensions 256 (circles), 512 (stars), and 1024 (x). 
For weak perturbations below the FGR regime the fidelity barely decays. In the
limit of strong perturbation $F_{\infty}$ saturates at $2/N$ (solid line). 
For intermediate values of $\delta$, $F_{\infty}$ is well approximated by the 
estimate of equation (9) with the proportionality constant $C_{CUE} = 3.6$.  
$F_{\infty}$ is obtained by averaging over 2000 map iterations starting 
at iteration $n = 2000$, well after the initial exponential decay. This is 
averaged over all $N$ initial eigenstates. The lower inset shows the 
initial exponential fidelity decay of the CUE map with $N = 1024$ averaged 
over all 1024 system eigenstates. The fidelity decay is plotted versus 
$\delta^2 n$ so that the exponential decay rates overlap and the saturation
level is easily seen. The perturbation strengths used are 0.1, 0.2, 0.3, 
0.4 (top to bottom). The upper inset shows a semi-log plot of the local 
density of states for a CUE map perturbed by a collective bit $z$-rotation, 
$\delta = 0.1,0.2,0.3$ and $0.4$ (bottom to top). The solid line is a 
Lorentzian of width $\Gamma = \overline{V_{mn}^2}/\Delta$ with 
$\overline{V_{mn}^2}$ determined numerically from the CUE map.}
\end{figure}

A similar analysis was carried out for random circular 
orthogonal ensemble (COE) maps. 
Random COE matrices can be created from CUE matrices, 
$COE = CUE*transpose(CUE)$ \cite{Haake}. Like 
the CUE maps, the COE maps have no classical analog and we introduce them here
as models for the behavior of quantum chaotic maps with COE eigenvector 
statistics and energy level spacings. 
Figure 3 shows $F_{\infty}$ versus perturbation strength for COE maps. 
Again, an approximate quadratic relationship emerges but with a 
different proportionality coefficient, $C_{COE} = 5.4$. 

The difference in proportionality constants is in line with the work of 
Prosen \cite{Prosen} who, using a random matrix theory 
argument, predicts a ratio of $3/2$ for 
$F_{\infty}^{COE}/F_{\infty}^{CUE}$ in the limit of strong perturbation.
We observe that this ratio holds for all perturbation strengths in 
the FGR regime. The calculated numerical average of 
$F_{\infty}^{COE}/F_{\infty}^{CUE}$ for the three Hilbert space
dimensions explored with perturbations in the FGR regime is $1.48 \simeq 3/2$.

We next study $F_{\infty}$ for a quantum system with a well defined 
classical analog, the quantum kicked top (QKT) \cite{qkt}\cite{Haake}. 
The QKT is an exemplary 
model of quantum chaos and has been used in previous studies of fidelity 
decay \cite{P1}\cite{Prosen}\cite{Jacq}. The QKT is a unitary map 
$U_{QKT} = exp(-i\pi J_y/2)exp(-ik J_z^2/j)$ acting on a Hilbert space of 
dimension $N = 2j+1$. $\vec{J}$ is the angular momentum operator in the 
irreducible representation and $k$ is the kick strength. A kick strength of 
$k = 12$ is used which is well in the chaotic region of the QKT. Since the
QKT shows anti-unitary symmetry, it is part of the COE class. The QKT has 
COE-like nearest neighbor level spacings 
\cite{Haake} and eigenvector statistics \cite{Zyc}. The same
perturbation, the collective $z$-rotation, is used. 

\begin{figure}
\epsfig{file=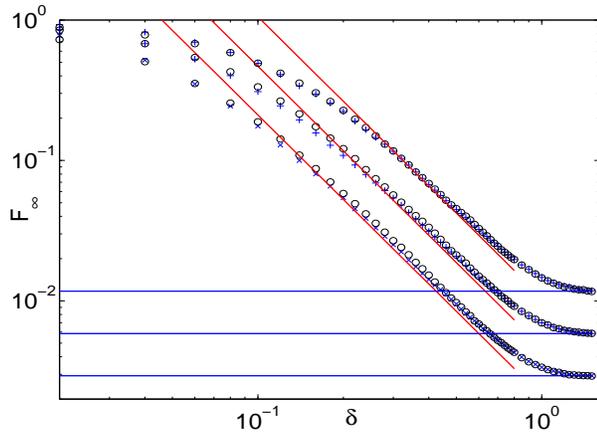, height=5.8cm, width=8cm}
\caption{ $F_{\infty}$ versus perturbation strength for initial eigenstates
of a random COE map (x) and the QKT with k = 12 (circles) of dimensions 
256, 512, and 1024 (from top to bottom). 
For weak perturbations below the FGR regime the fidelity barely decays. In the
limit of strong perturbation $F_{\infty}$ saturates at $3/N$ (solid line). 
For intermediate values of $\delta$, $F_{\infty}$ is well approximated by the 
estimate of equation (9) with the proportionality constant $C_{COE} = 5.4$.  
The numerical value of $F_{\infty}$ is determined in the same manner as for 
the CUE maps.}
\end{figure}

It should be noted that the data
for the QKT and COE maps are very similar. This is expected in that, as has
been conjectured and demonstrated 
in a number of works, quantum chaotic systems have 
statistical \cite{BGS}\cite{Zyc} and dynamic features \cite{J} 
similar to those of the canonical random matrix theory ensembles.  

The QKT is a system with a classical analog and has symmetries not found
in random matrices. It is interesting to see what effect these symmetries,
or invariant subspaces have on $F_{\infty}$. To do this, $F_{\infty}$ is
calculated for the oe subspace (odd under $180^o$ rotations around the 
$y$-axis \cite{P1}) of the QKT which has dimension $N = j$. 
The results are shown in figure 3 and again we see that  
$F_{\infty}$ approximately follows a quadratic decrease with increased 
perturbation strength. However, while the saturation level at the limit of 
strong perturbation does reach the expected $3/N$ at the same perturbation
strength as for the full QKT, the intermediate perturbation strengths
lead to a saturation level that is higher then for the full QKT. The 
coefficient $C_{oe}$ is significantly higher than 
that of the CUE or COE maps. 

\begin{figure}
\epsfig{file=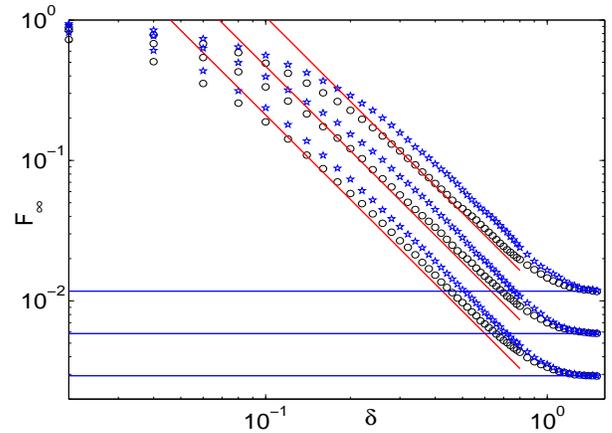, height=5.8cm, width=8cm}
\caption {$F_{\infty}$ versus perturbation strength for initial eigenstates
of the oe subspace of the QKT with $k = 12$ (stars) and the full QKT with
$k = 12$ (circles) for Hilbert space dimensions 256, 512 and 1024 (top to 
bottom). For weak perturbations below the FGR regime the 
fidelity barely decays. In the limit of strong perturbation 
$F_{\infty}$ saturates at $3/N$ (solid line). 
For intermediate values of $\delta$ of the full kicked top, 
$F_{\infty}$ is well approximated by the 
estimate of equation (9) with the proportionality constant $C_{COE} = 5.4$.  
However, $F_{\infty}$ for the oe subspace does not match $F_{\infty}$ of 
the full kicked top for these perturbation strengths. 
}
\end{figure}

In conclusion, we have given a theoretical argument estimating 
the saturation level of fidelity decay, $F_{\infty}$, for initial states that
are eigenstates of the system for intermediate perturbation strengths. 
Numerical simulations for systems with and
without classical analogs agree with the theoretical predictions. However, 
the presence of invariant subspaces appears to influence 
the saturation level of the fidelity decay.
 
This work was supported by DARPA/MTO through ARO grant DAAG55-97-1-0342 
and by the Cambridge-MIT Institute.


\begin{thebibliography}{99}
\bibitem{Berry}
M.V. Berry and M. Tabor, Proc. Roy. Soc. Lond. {\bf A356}, 375 (1977).
\bibitem{BGS}
O. Bohigas, M.J. Giannoni, C. Schmit, Phys. Rev. Lett. {\bf 52}, 1, 1984.
\bibitem{Haake}
F. Haake, {\it Quantum Signatures of Chaos} (Springer, New York, 1991).
\bibitem{P1}
A. Peres, Phys. Rev. A {\bf 30}, 1610 (1984); {\it Quantum Theory: Concepts 
and Methods}, Kluwer Academic Publishers (1995). 
\bibitem{SC}
R. Schack, C. M. Caves, Phys. Rev. Lett. {\bf 71} 525-528, 1993. 
\bibitem{ZP1}
W.H. Zurek, J.P. Paz, Physica D, {\bf 83}, 300-308, 1995.
\bibitem{Prosen}
T. Prosen, M. Znidaric, J. Phys. A {\bf 35} 1455, 2002.
\bibitem{Jacq}
Ph. Jacquod, P.G. Silvestrov, C.W.J. Beenakker, Phys. Rev. E {\bf 64}, 055203
(2001)
\bibitem{J}
J. Emerson, Y.S. Weinstein, S. Lloyd, D.G. Cory, quant-ph/0207099
\bibitem{Tom}
N. Cerruti and S. Tomsovic, Phys. Rev. Lett. {\bf 88}, 054103 (2002).
\bibitem{Jala}
R.A. Jalabert and H.M. Pastawski, Phys. Rev. Lett. {\bf 86}, 2490 (2001); F.
Cucchietti, C.H. Lewenkopf, E.R. Mucciolo, H. Pastawski, R.O. Vallejos 
nlin.CD/0112015.
\bibitem{Cas}
G. Beneti and G. Casati, quant-ph/0112060.
\bibitem{Wigner}
E.P. Wigner, Ann. Math. {\bf 62}, 548 (1955); {\bf 65}, 203 (1957).
\bibitem{FCIC}
Y.V. Fyodorov, O.A. Chubykalo, F.M. Izrailev, and G. Casati, Phys.~Rev.~
Lett.~{\bf 76}, 1603 (1996).
\bibitem{Jacq2}
Ph. Jacquod and D.L. Shepelyansky, Phys.~Rev.~Lett.~{\bf 75}, 3501 (1995).
\bibitem{Jacq3}
 Ph. Jacquod, I. Adagideli, C.W.J. Beenakker, accepted for publication in 
Phys. Rev. Lett.
\bibitem{Viola}
L. Viola, E. Fortunato, M.A. Pravia, E. Knill, R. Laflamme, D.G. Cory, Science
{\bf 293}, 2059 (2001).
\bibitem{qkt}
F. Haake, M. Kus, R. Scharf, Z. Phys. B, {\bf 65}, 381 (1987).
\bibitem{Zyc}
F. Haake, K. Zyczkowski, Phys. Rev. A {\bf 42}, 1013 (1990).


\end{thebibliography}
\end{document}